\documentclass[10pt]{revtex4}
\usepackage{hyperref}
\usepackage{amsmath}
\usepackage[psamsfonts]{amssymb }
\usepackage{mathrsfs}

\begin{document}

\title{\Large BRST Symmetry for Torus Knot }

 \author{Vipul Kumar Pandey\footnote {e-mail address: vipulvaranasi@gmail.com}}
\author{ Bhabani Prasad Mandal\footnote {e-mail address: bhabani.mandal@gmail.com}}

\affiliation { Department of Physics, 
Banaras Hindu University, 
Varanasi-221005, INDIA.  }

\begin{abstract}
We develop BRST symmetry for the first time for a particle on the surface of a torus knot by analyzing the constraints of the system. The theory contains $2^{nd}$ class constraints and has been extended by introducing Wess-Zumino term to the convert it into a theory with first class constraints. BFV analysis of the extended theory is performed to construct BRST/anti-BRST symmetries for the particle on torus knot. The nilpotent BRST/anti-BRST charges which generate such symmetries are constructed explicitly. The states annihilated by these nilpotent charges  consist the physical Hilbert space. We indicate how various effective theories on the surface of the torus knot are related through the generalized version of BRST transformation with finite field dependent parameters.

\end{abstract}
\maketitle
\section{Introduction}\label{Introduction} 
Knot \cite{1,2} theory, based on mathematical concepts has found immense  applications in various branch of frontier physics. Knot invariants in physical systems were introduced long ago and has got considerable impact during last one and half decades \cite{1,2,3,4,5,6,7,8,9}, especially when interpreted as Wilson loop observable in Chern-Simon(CS) theory \cite{7}. The discussion on topological string approach to the torus knot invariants are presented in Ref. \cite{7}. In the context of gauge theory, knot invariant theories relate 3d symmetry on the CS sub manifolds and 3d SUSY gauge theory. It also plays important role in various other problems like, inequivalent quantization problem \cite{4}, in the role of topology in defining vacuum state in gauge theories\cite{5}, in understanding band theory in solid \cite{6}. Various longstanding problems related with  connection between knot theory and quantum field theory were discussed in \cite{7}. Dynamics and symmetries of the particles constraints to move on the surface of a torus knot has recently be addressed through Hamiltonian analysis \cite{9}

BRST quantization \cite{11} is an important and powerful technique to deal with system with constraints \cite{10}. It enlarges the phase space of a gauge theory and restore the symmetry of the gauge fixed action in the extended phase space keeping the physical contents of the theory unchanged. BRST symmetry plays a very important role in renormalizing spontaneously broken theories, like standard model and hence is extremely important to investigate it for different systems. To the best of our knowledge BRST formulation for the particle on torus knot has not been developed yet. This motivates us in the study of BRST symmetry for a particle on the surface of the torus knot. In the present work we make an important step forward in formulating BRST symmetry for a particle constrained to move on torus knot. We study the particle on a torus knot following the technique of Dirac's constraints analysis \cite{10}. The system is shown to contain $2^{nd}$ class constraints. We introduce Wess-Zumino term to recast the system in a gauge invariant fashion in the extended Hilbert space. We further develop BFV (Batalin - Fradkin - Vilkovisky) formulation of this extended theory using the constraints in the theory            \cite{12,13,14,15}. The nilpotent BRST and anti-BRST charges are constructed which generate the transformations using the constraints in the theory.  These nilpotent BRST charges annihilate the states in the physical Hilbert space which is shown to be consistent with the constraints present in the theory. We indicate how various BRST invariant effective theories on the surface of a torus are interlinked by considering finite field dependent version of the BRST (FFBRST) transformation, introduced by Joglekar and Mandal \cite{16} about 22 years ago. FFBRST transformations are the generalization of usual BRST transformation where the usual infinitesimal, anti-commuting constant transformation parameter is replaced by field dependent but global and anti-commuting parameter. Such generalized transformation protects the nilpotency and retains the symmetry of the gauge fixed effective actions. The remarkable property of such transformations are that they relates the generating functional corresponding to different effective actions. The non-trivial Jacobian of the path integral measure under such a finite transformation is responsible for all the new results.
In the virtue of this remarkable property FFBRST transformation have been investigated extensively  and has found many application in various gauge field theoretic systems.\cite{17,18,19,20,21,22,23,24,25,26,27,272,33,34,35,36}. Similar generalization of BRST transformation with same motivation and goal has also been carried more recently in slightly different
manner where Jacobian for such transformation is calculated without using any ansatz \cite{275,28,29,30,31,32}.
                   
                    We  now present the plan of the paper. In Sec II we analyze the constraints of the system. The theory of a particle on torus knot has been extended by introducing WZ term in sec III. BFV formulation is presented in sec IV. Nilpotent charges has been constructed in sec V. In sec VI FFBRST for torus knot is presented. Sec VII is kept for concluding remarks. 
\section{Particle on a Torus Knot }\label{Particle on a Torus Knot }
In knot theory, a torus knot is a special kind of knot that lies on the surface of un-knotted torus in $R^3$. It is specified by a set of co-prime integers $p$ and $q$. A torus knot of type $(p, q)$ winds p times around the rotational symmetry axis of the torus and q times around a circle in the interior of the torus. The toroidal co-ordinate system is a suitable choice to study this system. Toroidal co-ordinates are related to Cartesian co-ordinates ($x_1, x_2, x_3$) in following ways
\begin{eqnarray}
x_1 = \frac{a \sinh\eta \cos\phi}{\cosh\eta - \cos\theta},\quad x_2 = \frac{a \sinh\eta \sin\phi}{\cosh\eta - \cos\theta}, \quad x_3 = \frac{a \sin\theta}{\cosh\eta - \cos\theta}
\label{ccoo}
\end{eqnarray}
where, $0\leq\eta\leq\infty$, $-\pi\leq\theta\leq\pi$ and $0\leq\phi\leq 2\pi$. A toroidal surface is represnted by some specific value of $\eta$ (say $\eta_0$). Parameters a and $\eta_0$ are written as $a^2 = R^2 - d^2$ and $\cosh\eta_0 = \frac{R}{D}$ where $R$ and $D$ are major and minor radius of torus respectively.

Similarly, toroidal coordinates can be represented in the form of Cartesian co-ordinates as,
\begin{eqnarray}
\eta = \ln\frac{d_1}{d_2},\quad \cos\theta = \frac {r^2 - a^2}{((r^2 - a^2)^2 + 4 a^2 {x_3}^2)^{\frac{1}{2}}}, \quad \phi = \tan^{-1}\frac{x_2}{x_1}
\label{tcoo}
\end{eqnarray}
where 
\begin{eqnarray}
d^2_1 = (\sqrt{x^2_1 + x^2_2} + a)^2 + x^2_3,\quad d^2_2 = (\sqrt{x^2_1 + x^2_2} - a)^2 + x^2_3, \quad r^2 = x^2_1 + x^2_2
\label{rct}
\end{eqnarray}
  $r$ is cylindrical radius.

Lagrangian for a particle residing on the surface of torus knot is given by  \cite{3,9}           
\begin{eqnarray}
L = \frac{1}{2} m a^2\frac{{\dot \eta}^2+ {\dot \theta}^2+ {\sinh\eta}^2{\dot\phi}^2}{(\cosh\eta - \cos\theta)^2} - \lambda (p\theta + q\phi)
\label{lpstk}
\end{eqnarray}
where $(\eta,\theta,\phi)$ are the toroidal coordinates for toric geometry. 
 
Constraint that forces the particle to move in knot is imposed as
\begin{eqnarray} 
\Omega_1 = (p\theta + q\phi)\approx 0
\label{pmc}
\end{eqnarray}
The Hamiltonian corresponding to this Lagrangian is then written as`
\begin{eqnarray}
H = \frac {(\cosh\eta - \cos\theta)^2}{m a^2}[{p^2_\eta} + {p^2 _\theta}+ \frac{{p^2_\phi}}{{\sinh^2\eta}}] + \lambda (p\theta + q\phi)
\label{hpstk}
\end{eqnarray}
Here ${p_\eta}$, ${p_\theta}$ and ${p_\phi}$ are canonical momenta corresponding to the co-ordinate $\eta$, $\theta$ and $\phi$.

Time evolution of the constraint $\Omega_1$ gives additional secondary constraint
\begin{eqnarray} 
\Omega_2 = \frac {(\cosh\eta - \cos\theta)^2}{m a^2}[{p p_\theta + \frac{q p_\phi}{\sinh^2\eta}}]\approx 0 
\label{secc}
\end{eqnarray}
Constraints $\Omega_1$ and $\Omega_2$ form second-class constraint algebra
\begin{eqnarray} 
\Delta_{kk'}(x,y) = \{\Omega_k (x), \Omega_{k'} (y)\} = \epsilon^{k k'} \frac {(\cosh\eta - \cos\theta)^2}{m a^2}[{p^2 + \frac{q^2}{\sinh^2\eta}}]\delta( x-y )
\label{copsc}
\end{eqnarray}
with $\epsilon^{12} = - \epsilon^{21} = 1$. In the next section we will convert this gauge variant theory to the gauge invariant theory in an extended Hilbert space.

\section{Wess - Zumino term and Hamiltonian Formulation}\label{Wess - Zumino term and Hamiltonian Formulation}
To construct a gauge invariant theory corresponding to this a gauge non-invariant model of particle on a torus knot, we introduce Wess - Zumino term in Lagrangian in Eq. \ref{pstk}.
For this purpose we enlarge the Hilbert space of the theory by introducing a new co-ordinate $\alpha$, called as Wess-Zumino term, through the redefinition of co-ordinates $\theta$, $\phi$ and $\lambda$ in the Lagrangian $L$ in Eq. \ref{lpstk} as follows
\begin{eqnarray}
\theta \rightarrow \theta -\frac{\alpha}{2p}, \quad\phi \rightarrow \phi - \frac{\alpha}{2q}, \quad\lambda \rightarrow \lambda + \dot\alpha 
\label{rdtpl}
\end{eqnarray}

With this redefinition of co-ordinates, modified Lagrangian is written as
\begin{eqnarray}
 L^{I} &=& \frac{1}{2} \frac{m a^2}{(\cosh\eta - \cos (\theta - \frac{\alpha}{2p}))^2}[{\dot \eta}^2+ {\dot \theta}^2+ {\sinh^2\eta}{\dot\phi}^2 + \frac{{\dot\alpha}^2}{4}(\frac{1}{p^2} + \frac{{\sinh^2\eta}}{q^2}) - \dot\alpha (\frac{\dot\theta}{p} + {\sinh^2\eta}\frac{\dot\phi}{q})] \nonumber\\ & - & (\lambda + \dot\alpha) (p\theta + q\phi - \alpha )
\label{rdld}
\end{eqnarray} 
which is invariant under following time-dependent gauge transformations
\begin{eqnarray} 
\delta \lambda =  {\dot f(t)}, \quad \delta \theta = -\frac{f(t)}{2p}, \quad\delta \phi = -\frac {f(t)}{2q}, \quad\delta \alpha = - f(t)\nonumber\\
\delta p_\eta = \delta p_\theta = \delta p_\phi  = 0, \quad \delta b = \delta \Pi_\alpha  = \delta \eta = \delta \Pi_\lambda = 0
\label{tdgtr}
\end{eqnarray}
where $f(t)$ is an arbitrary function of time.  To construct the Hamiltonian for this gauge invariant theory we construct the
canonical momenta corresponding to this modified Lagrangian  and are  written as
\begin{eqnarray}
p_\eta &=&  \frac{m a^2}{(\cosh\eta - \cos (\theta - \frac{\alpha}{2p}))^2}\dot\eta, \quad \Pi_\lambda = 0 \nonumber\\
p_\theta &=& \frac{m a^2}{(\cosh\eta - \cos (\theta - \frac{\alpha}{2p}))^2}(\dot\theta - \frac{\dot\alpha}{2p})\nonumber\\
p_\phi &=& \frac{m a^2 \sinh^2\eta}{(\cosh\eta - \cos (\theta - \frac{\alpha}{2p}))^2}(\dot\phi - \frac{\dot\alpha}{2q})\nonumber\\
\Pi_\alpha &=& \frac{m a^2}{2(\cosh\eta - \cos (\theta - \frac{\alpha}{2p}))^2}\{\frac{\dot\alpha}{2}(\frac{1}{p^2} + \frac{{\sinh^2\eta}}{q^2}) + (\frac{\dot\theta}{p} + {\sinh^2\eta}\frac{\dot\phi}{q})\} - (p\theta + q\phi - \alpha ) 
\label{cmmld}
\end{eqnarray} 
The only primary constraint for this extended theory is
\begin{eqnarray}
\Psi_1 \equiv \Pi_\lambda \approx 0
\label{mpmc}
\end{eqnarray} 
The Hamiltonian corresponding to Lagrangian $L^{I}$ is written as
\begin{eqnarray}
 H^{I} = p_\eta \dot\eta + p_\theta \dot\theta + p_\phi \dot\phi + \Pi_\alpha \dot\alpha -  L^{I}
\label{hdcld} 
\end{eqnarray}
The total Hamiltonian after using Lagrange multiplier $\beta$ corresponding to the primary constraint $\Pi_\lambda$ is obtained as 
\begin{eqnarray}
H^{I}_{T} = \frac{(\cosh\eta - \cos (\theta - \frac{\alpha}{2p}))^2}{2m a^2}[p^2_\eta + p^2_\theta + \frac {p^2_\phi}{\sinh^2\eta}] - \lambda (\Pi_\alpha + \frac{p_\theta}{2p} + \frac{p_\phi}{2q}) + \beta \Pi_\lambda
\label{thd}
\end{eqnarray}
Using Dirac's method of constraint analysis \cite{10}, we obtain secondary constraint
\begin{eqnarray}
\Psi_2 \equiv (\Pi_\alpha + \frac{p_\theta}{2p} + \frac{p_\phi}{2q}) \approx 0
\label{mscc}
\end{eqnarray}
There is no  tertiary constraint corresponding to this total Hamiltonian as
\begin{eqnarray}
\Psi_3 = \dot\Psi_2  = [ H^{I}_{T},(\Pi_\alpha + \frac{p_\theta}{2p} + \frac{p_\phi}{2q})] = 0
\label{trc}
\end{eqnarray}
This extended theory thus has only first class constraints.

\section{BFV Formalism for Torus Knot}\label{BFV Formalism for Torus Knot}
To discuss all possible nilpotent symmetries we further extend the theory using BFV formalism \cite{12,13,14,15}. In the BFV formulation associated with this system, we introduce a pair of canonically conjugate ghost fields (c,p) with ghost number 1 and -1 respectively, for the primary constraint $\Pi_\lambda \approx 0$ and another pair of ghost fields $(\bar c, \bar p)$ with ghost number -1 and 1 respectively, for the secondary constraint, $(\Pi_\alpha + \frac{p_\theta}{2p} + \frac{p_\phi}{2q}) \approx 0$.
The effective action for a particle on surface of the torus knot in extended phase space is then written as
\begin{eqnarray}
S_{eff}& = &\int dt \big[ p_{\eta}\dot\eta + p_{\theta}\dot\theta + p_{\phi}\dot\phi + \Pi_\alpha \dot\alpha - \Pi_{\lambda}\dot\lambda
-\frac{(\cosh\eta - \cos (\theta - \frac{\alpha}{2p}))^2}{2m a^2}\{p^2_\eta + p^2_\theta\nonumber\\ & + &\frac{p^2_\phi}{\sinh^2\eta}\}+{\dot c} P + \dot{\bar c} \bar P - \{Q_b,\psi\} \big]
\label{eapstk}
\end{eqnarray}

Where $Q_b$ is BRST charge and has been  constructed using the constraints of the system as 
\begin{eqnarray}
Q_b = ic(\Pi_\alpha + \frac{p_\theta}{2p} + \frac{p_\phi}{2q} )-i\bar P \Pi_\lambda
\label{brch}
\end{eqnarray}  
The canonical brackets for all dynamical variables are written as
\begin{eqnarray}
 [\eta, p_\eta] = [\theta, p_\theta] = [\phi, p_\phi] = [\alpha, \Pi_\alpha] = [\lambda, \Pi_\lambda] =  \{\bar c\, \dot c\} = i; \quad\{c, \dot{\bar c}\} = - i
\label{cbfadb}
\end{eqnarray} 
Nilpotent BRST transformation corresponding to this action is constructed using the relation $s_b\Phi = - [Q_b, \Phi]_{\pm}$ which is related to infinitesimal BRST transformation as $\delta_b \Phi = s_b \Phi \delta \Lambda$. Here $\delta \Lambda$ is infinitesimal BRST parameter. Here $-$ sign is for bosonic and $+$ is for fermionic variable. The BRST transformation for the particle on a torus knot is then  written as
\begin{eqnarray} 
s_b \lambda =  {\bar P}, \quad s_b \theta = -\frac{c}{2p}, \quad s_b \phi = -\frac {c}{2q}, \quad s_b \alpha = - c\nonumber\\
s_b p_\eta = s_b p_\theta = s_b p_\phi  = 0, \quad s_b P = (\Pi_\alpha + \frac{p_\theta}{2p} + \frac{p_\phi}{2q} )\nonumber\\
s_b \bar c = \Pi_\lambda = b,\quad s_b c = s_b b = s_b \Pi_\alpha  = s_b \eta = s_b  \Pi_\lambda = 0
\label{brtrf}
\end{eqnarray},
One can check that these transformations are nilpotent.

In BFV formulation the generating functional is independent of gauge fixing fermion \cite{12,13,14}, hence we have liberty to choose it in the convenient form as
\begin{eqnarray}
\Psi = p\lambda + \bar c (p\theta + q\phi + \alpha + \frac{\Pi_\lambda}{2})
\label{psi} 
\end{eqnarray}
Using the expressions for $Q_b$ and $\Psi$, Effective action (\ref{eapstk}) is written as
\begin{eqnarray}
S_{eff}& = &\int dt \big[ p_\eta \dot\eta + p_\theta\dot\theta + p_\phi\dot\phi + \Pi_\alpha \dot\alpha - \Pi_\lambda\dot\lambda - \frac{(\cosh\eta - \cos (\theta - \frac{\alpha}{2p}))^2}{2m a^2}\{p^2_\eta + p^2_\theta\nonumber\\ & + &\frac{p^2_\phi}{\sinh^2\eta}\}+{\dot c} P + \dot{\bar c} \bar P - \bar P P + \lambda (\Pi_\alpha + \frac{p_\theta}{2p} + \frac{p_\phi}{2q}) + 2c\bar c + \Pi_\lambda (p\theta + q\phi + \alpha + \frac{\Pi_\lambda}{2}) \big]
\label{eleps}
\end{eqnarray}
and the generating functional for this effective theory is represented as 
\begin{eqnarray}
Z_\psi &=& \int D \phi \quad  exp \big[iS_{eff} \big]
\label{gnf}  
\end{eqnarray}
The measure $ D\phi =\prod_i d\xi_i $ , where $\xi_i$ are all dynamical variables of the theory.
Now integrating this generating functional over P and $\bar P$, we get 
\begin{eqnarray}
{Z_\psi} & = & \int D \phi' exp \big[i\int dt \big[ p_\eta \dot\eta + p_\theta\dot\theta + p_\phi\dot\phi + \Pi_\alpha \dot\alpha - \Pi_\lambda\dot\lambda - \frac{(\cosh\eta - \cos (\theta - \frac{\alpha}{2p}))^2}{2m a^2}\{p^2_\eta + p^2_\theta\nonumber\\ & + &\frac{p^2_\phi}{\sinh^2\eta}\} + \dot{\bar c}{\dot c} + \lambda (\Pi_\alpha + \frac{p_\theta}{2p} + \frac{p_\phi}{2q}) - 2\bar c c + \Pi_\lambda (p\theta + q\phi + \alpha + \frac{\Pi_\lambda}{2}) \big] \big]
\label{gfaioppb}    
\end{eqnarray}
where $D\phi'$ is the path integral measure for effective theory when integrations over fields P and $\bar P$ are carried out.
Further integrating over $\Pi_\lambda$ we obtain an effective generating functional as
\begin{eqnarray}
{Z_\psi} & = & \int D \phi'' exp \big[i\int dt \big[ p_\eta \dot\eta + p_\theta\dot\theta + p_\phi\dot\phi + \Pi_\alpha \dot\alpha - \frac{(\cosh\eta - \cos (\theta - \frac{\alpha}{2p}))^2}{2m a^2}\{p^2_\eta + p^2_\theta\nonumber\\ & + &\frac{p^2_\phi}{\sinh^2\eta}\} + \dot{\bar c}{\dot c} + \lambda (\Pi_\alpha + \frac{p_\theta}{2p} + \frac{p_\phi}{2q}) - 2\bar c c -  \frac{\{\dot\lambda - (p\theta + q\phi + \alpha)\}^2}{2} \big] \big]
\label{gfaiopl}    
\end{eqnarray}
where $D\Phi''$ is the path integral measure corresponding to all the dynamical variables involved in the effective action. The BRST symmetry transformation for this effective theory is written as 
\begin{eqnarray} 
 s_b \theta = -\frac{c}{2p}, \quad s_b \phi = -\frac {c}{2q}, \quad s_b \alpha = - c\nonumber\\
s_b p_\eta = s_b p_\theta = s_b p_\phi  = 0, \quad s_b \bar c = -\{\dot\lambda - (p\theta + q\phi + \alpha)\} \nonumber\\
s_b \lambda =  {\dot c},\quad s_b c = s_b b = s_b \Pi_\alpha  = s_b \eta = s_b \Pi_\lambda = 0
\label{btrf}
\end{eqnarray}

\section{BRST and Anti-BRST charge}\label{BRST and Anti-BRST charge} 
 
In this section we show that physical subspace of the system is consistent with the constraints of the system.
The physical states are annihilated by the BRST charge in Eq. \ref{brch}
\begin{eqnarray}
Q_b|\psi\rangle =0= \{ i c (\Pi_\alpha + \frac{p_\theta}{2p} + \frac{p_\phi}{2q}  ) - i \dot c \Pi_\lambda\} |\psi\rangle  =  i c ( \Pi_\alpha + \frac{p_\theta}{2p} + \frac{p_\phi}{2q} )|\psi\rangle  - i \dot c \Pi_\lambda\} |\psi\rangle 
\label{bcos}
\end{eqnarray}

This implies that
\begin{eqnarray}
  ( \Pi_\alpha + \frac{p_\theta}{2p} + \frac{p_\phi}{2q} )|\psi\rangle  = 0, \quad \Pi_\lambda |\psi\rangle  = 0
\label{bcost}
\end{eqnarray}
The Hamiltonian(\ref{thd}) is  also invariant under anti-BRST transformation in which role of $c$ and $-\bar c$ are interchanged. 
  Anti-BRST transformations for this theory are written as  
\begin{eqnarray} 
\bar s_b \theta = \frac{\bar c}{2p}, \quad\bar s_b \phi = \frac {\bar c}{2q}, \quad\bar s_b \alpha =  \bar c\nonumber\\
\bar s_b p_\eta = \bar s_b p_\theta = \bar s_b p_\phi  = 0, \quad\bar s_b  c = \{\dot\lambda - (p\theta + q\phi + \alpha)\} \nonumber\\
\bar s_b \lambda = - {\dot {\bar c}},\quad \bar s_b c = \bar s_b b = \bar s_b \Pi_\alpha  = \bar s_b \eta = \bar s_b \Pi_\lambda = 0
\label{abtrf}
\end{eqnarray}
The nilpotent charge for the anti-BRST symmetry  in (\ref{abtrf})  is constructed as
\begin{eqnarray}
Q_{ab}  &= &- i \bar c (\Pi_\alpha + \frac{p_\theta}{2p} + \frac{p_\phi}{2q} ) + i \dot{\bar c} \Pi_\lambda
\label{abchrg}
\end{eqnarray}
Like BRST charge, anti-BRST charges $Q_{ab}$  also generates the anti-BRST transformations in (\ref{abtrf}) through the following commutation and anti-commutation relations

\begin{eqnarray}
s_{ab}\theta=\ -[Q_{ab}, \theta] &=& \frac{\bar c}{2p} \nonumber\\
s_{ab}\phi=\ -[Q_{ab}, \phi ] &=& \frac{\bar c}{2q} \nonumber\\
s_{ab}\alpha=\ -[Q_{ab}, \alpha ] &=&  \bar{c} \nonumber\\
s_{ab}\lambda =\ -[Q_{ab}, \lambda] &=& - \dot{\bar c} \nonumber\\
s_{ab} c= - \ \{Q_{ab,}  c\} &=&  - \Pi_\lambda
\label{abcacr}
\end{eqnarray}

Anti-BRST charge too annihilates the states of physical Hilbert space. 
\begin{eqnarray}
Q_{ab}|\psi\rangle  &=& 0 \nonumber\\
-i \bar c (\Pi_\alpha + \frac{p_\theta}{2p} + \frac{p_\phi}{2q}  ) + i \dot{\bar c} \Pi_\lambda|\psi\rangle  &=&0 
\label{abos}
\end{eqnarray}
or
\begin{eqnarray} 
( \Pi_\alpha + \frac{p_\theta}{2p} + \frac{p_\phi}{2q}  )|\psi\rangle = 0, \quad \Pi_\lambda|\psi\rangle  = 0   
\label{abas}
\end{eqnarray} 
 Anti-BRST charge too project on the physical subspace of total Hilbert space. Thus anti-BRST charge plays exactly  same role as BRST charge.
It is straight forward to check  that these charges are nilpotent i.e. $ Q_b^2 =0= Q_{ab}^2 $ and satisfy
\begin{equation}
\{Q_b,\ \ Q_{ab} \}=0
\label{babcr}
\end{equation}

\section{FFBRST for Torus Knot}\label{FFBRST for Torus Knot}
In this section we show that these nilpotent symmetries can be generalized by making the parameter finite and field dependent following the work of Joglekar and Mandal [17]. The BRST transformations  are generated from BRST charge using relation $\delta_b \phi = -[Q,\phi]\delta\Lambda$ where $\delta\Lambda$ is infinitesimal anti-commuting global parameter. Following their technique  the anti-commuting BRST parameter $\delta\Lambda$ is generalized to be finite-field dependent instead of infinitesimal but space time independent parameter $\Theta[\phi]$. Since the parameter is finite in nature unlike the usual case  the path integral measure is not invariant under such finite transformation.  The Jacobian for these transformations for certain $\Theta[\phi]$ can be calculated by following way.  
\begin{eqnarray}
D\phi &=& J(k)D\phi'(k)\nonumber\\
       &=& J(k+dk)D\phi'(k+dk)
\label{pimr}
\end{eqnarray}
where  a  numerical parameter $k$ ($0\leq k \leq 1$), has been introduced  to execute the finite transformation in a mathematically convenient way. All the fields are taken to be $k$ dependent in such a fashion that $\phi(x,0) = \phi(x)$ and $\phi(x,k = 1) = \phi'(x)$. 
$S_{eff}$ is invariant under FFBRST which is constructed by considering successive infinitesimal BRST transformation $(\phi(k)\rightarrow\phi(k+dk))$. The nontrivial Jacobian $J(k)$ can be written as local functional of fields and will be replaced as $e^{iS_1[\phi(k),k]}$ if the condition \cite{17}

\begin{eqnarray}
\int D\phi(k)\big[\frac{1}{J(k)}\frac{d J(k)}{d k}-i\frac{dS_1}{dk}\big] e^{i (S_1 + S_{eff})} = 0 
\label{ffbc}
\end{eqnarray}
holds. Where $\frac{dS_1}{dk}$ is a total derivative of $S_1$ with respect to k in which dependence on $\phi(k)$ is also differentiated. The change in Jacobian is calculated as 
\begin{eqnarray}
\frac{J(k)}{J(k+dk)} &=& \Sigma_{\phi}{\pm}\frac{\delta \phi(x,k+dk)}{\delta \phi(x,k)}\nonumber\\
                     &=& \frac{1}{J(k)}\frac{d J(k)}{d k} d k
 \label{jacbc}
\end{eqnarray}
${\pm}$ is for bosonic and fermionic fields ($\phi$) respectively.
We know that the effective action for a particle on surface of torus knot using BFV formulation is written in (\ref{eapstk}) and the BRST transformation is given by (\ref{brtrf}). The finite version of this BRST is then written as
\begin{eqnarray}
\delta_b \lambda =  {\bar P}\Theta, \quad \delta_b \theta = -\frac{c}{2p}\Theta, \quad\delta_b \phi = -\frac {c}{2q}\Theta, \quad\delta_b \alpha = - c\Theta\nonumber\\
\delta_b p_\eta = \delta_b p_\theta = \delta_b p_\phi  = 0, \quad\delta_b P = (\Pi_\alpha + \frac{p_\theta}{2p} + \frac{p_\phi}{2q} )\Theta\nonumber\\
\delta_b \bar c = \Pi_\lambda\Theta = b\Theta,\quad \delta_b c = \delta_b b = \delta_b \Pi_\alpha  = \delta_b \eta = \delta \Pi_\lambda = 0
\label{ffbt}
\end{eqnarray}
where $\Theta$ is finite field dependent, global and anti-commuting parameter. It is straight forward to check that under this transformation too, effective action in (\ref{eleps}) is invariant. 
 Generating functional for this effective theory is  then written as
\begin{eqnarray}
Z_{\psi} &=&\int D \Phi exp [i\int dt \big[p_\eta \dot\eta + p_\theta\dot\theta + p_\phi\dot\phi + \Pi_\alpha \dot\alpha - \Pi_\lambda\dot\lambda - \frac{(\cosh\eta - \cos (\theta - \frac{\alpha}{2p}))^2}{2m a^2}\{p^2_\eta + p^2_\theta\nonumber\\ & + &\frac{p^2_\phi}{\sinh^2\eta}\} + {\dot c} P + \dot{\bar c} \bar P - \bar P P + \lambda (\Pi_\alpha + \frac{p_\theta}{2p} + \frac{p_\phi}{2q}) - 2\bar c c + \Pi_\lambda (p\theta + q\phi + \alpha + \frac{\Pi_\lambda}{2}) \big]\big ] ]
\label{gfet} 
\end{eqnarray}
where,
\begin{eqnarray}
D\Phi = \prod d\eta dp_\eta d\theta dp_\theta d\phi dp_\phi  d\lambda d\Pi_\lambda dP d\bar P dc d\bar c 
\label{pimrps}
\end{eqnarray} 
is the path integral measure in the total phase space. This path integral measure is not invariant under such FFBRST transformation as already mentioned. It gives rise to a Jacobian in the extended phase space which is  calculated using (\ref{jacbc}). 
Using the condition in (\ref{ffbc}), one can calculate the extra part in the action $S_1$ for some specific choices of the finite parameter $\Theta$.

Now we consider a simple example of FFBRST to show the connection between two effective theories explicitly.
For  that we choose  finite BRST parameter $\Theta = \int  dk \Theta^\prime (k) $ where $\Theta^\prime$ is given as
\begin{eqnarray}
\Theta^\prime= i \gamma\int d^4 y \bar c(y,k) \Pi_\lambda(y,k).
\label{thtp} 
\end{eqnarray} 
The change in Jacobian  is calculated for this particular parameter as,
\begin{eqnarray}
 \frac{1}{J(k)}\frac{d J(k)}{d k} = - i \gamma\int d^4 y  \Pi^2_\lambda (y,k)
 \label{cjcb}
\end{eqnarray}
We make an ansatz for $S_1$ as,
\begin{eqnarray}
S_1 = i\int d^4 x \xi_1(k) \Pi^2_\lambda
\label{ansso}
\end{eqnarray}
Where $\xi_1(k)$ is a $k$ dependent arbitrary parameter.  
Now,
\begin{eqnarray}
\frac{dS_1}{dk} = i\int d^4 x \xi'_1(k) \Pi^2_\lambda
\label{csowk}
\end{eqnarray}
By satisfying the condition in (\ref{ffbc}) we find $\xi_1 = - \gamma k$. The FFBRST with finite parameter $\Theta$ as given in (\ref{thtp})changes this generating functional as,
\begin{eqnarray}
Z &=& \int D\phi(k)e^{i (S_1 + S_{eff})}\nonumber\\
&=&\int D \Phi exp [i\int dt \big[ p_\eta \dot\eta + p_\theta\dot\theta + p_\phi\dot\phi + \Pi_\alpha \dot\alpha - \Pi_\lambda\dot\lambda - \frac{(\cosh\eta - \cos (\theta - \frac{\alpha}{2p}))^2}{2m a^2}\{p^2_\eta + p^2_\theta + \frac{p^2_\phi}{\sinh^2\eta}\} \nonumber\\ &+& {\dot c} P + \dot{\bar c} \bar P - \bar P P + \lambda (\Pi_\alpha + \frac{p_\theta}{2p} + \frac{p_\phi}{2q}) - 2\bar c c + \Pi_\lambda (p\theta + q\phi + \alpha) + (\frac{\lambda'}{2} - \gamma k) \frac{\Pi^2_\lambda}{2}) \big]\big ] 
\label{gf} 
\end{eqnarray} 
Here generating functional at $k=0$ will give pure theory for a free particle on a surface of torus with a gauge parameter $\lambda'$ and at $k=1$, the generating functional for same theory with a different gauge parameter $\lambda'' = \lambda' - 2 \gamma$. 
Even though we have considered a very simple example, our formulation is valid to connect any two generating functionals corresponding to different effective actions using FFBRST transformation with suitable parameter.
 
\section{Conclusion}\label{Conclusion}
 Mathematical concept of knot theory is very useful in describing various physical systems and it has been extensively used to study many different phenomena in physics. However there was no BRST formulation for particle on the surface of the torus knot. In this work we systematically developed the BRST/anti-BRST formulation for the first time for a particle moving on a torus knot. Using Dirac's constraint analysis we found all the constraints of this system. Further we have extended this theory to include Wess-Zumino term to recast this theory as gauge theory. Using BFV formulation BRST/Anti-BRST invariant effective action for a particle moving on a torus knot has been developed.  Nilpotent charges which generate these symmetries have been calculated explicitly. The physical states which are annihilated by these nilpotent charges are consistent  with the constraints of the system.  Our formulation is independent of particular choice of  a torus. We further have extended the BRST formulation by considering the transformation parameter finite and field dependent. We indicate how different effective theories on the surface of the torus knot are related through such a finite transformation through the non-trivial Jacobian factor. In support of our result we explicitly relate the generating functions of two effective theories with different gauge fixing parameters. Using FFBRST with suitable finite parameter the connection between any two effective theories can be made in a straight forward manner following the prescriptions outlined in this work.

                               One of us (VKP) acknowledges University Grant Commission(UGC), India  for its financial assistance under CSIR-UGC JRF/SRF scheme.

\end{document}